\documentclass[twocolumn,prl,aps,showpacs,amsmath,amssymb]{revtex4-1}
\usepackage{graphics}
\usepackage{graphicx}
\usepackage{dcolumn}
\usepackage{bm}
\usepackage{mathrsfs}
\usepackage{pstricks}
\usepackage{color}
\usepackage{braket}
\usepackage{slashed}
\usepackage{amsmath}
\usepackage{epsfig}
\usepackage{amsfonts}
\usepackage{amssymb}
\newcommand\Si{\textrm{Si}}

\newcommand\Ci{\textrm{Ci}}

\def\beq{\begin{equation}}
\def\eeq{\end{equation}}
\def\bea{\begin{eqnarray}}
\def\eea{\end{eqnarray}}
\def\nn{\nonumber}
\begin{document}

\title{The Fermi problem in disordered systems}
\author{G. Menezes}
\email{gabrielmenezes@ufrrj.br}
\affiliation{Grupo de F\'isica Te\'orica e Matem\'atica F\'isica, Departamento de F\'isica, Universidade Federal Rural do Rio de Janeiro, 23897-000 Serop\'edica, RJ, Brazil}
\author{N. F. Svaiter}
\email{nfuxsvai@cbpf.br}
\affiliation{Centro Brasileiro de Pesquisas F\'{\i}sicas, 22290-180 Rio de Janeiro, RJ, Brazil}
\author{H. R. de Mello}
\email{hrmello@if.ufrj.br}
\affiliation{Instituto de F\'isica, Universidade Federal do Rio de Janeiro, 21941-972 Rio de Janeiro, RJ, Brazil}
\author{C. A. D. Zarro}
\email{carlos.zarro@if.ufrj.br}
\affiliation{Instituto de F\'isica, Universidade Federal do Rio de Janeiro, 21941-972 Rio de Janeiro, RJ, Brazil}

\begin{abstract}
We revisit the Fermi two-atoms problem in the framework of disordered systems. In our model we consider a two-qubits system linearly coupled with a quantum massless scalar field. We analyze the energy transfer between the qubits under different experimental perspectives. In addition, we assume that the coefficients of the Klein-Gordon equation are random functions of the spatial coordinates. The disordered medium is modeled by a centered, stationary and Gaussian process. We demonstrate that the classical notion of causality emerges only in the wave zone in the presence of random fluctuations of the light cone. Possible repercussions are discussed.
\end{abstract}


\maketitle

{\it Introduction.}---Causality is the most essential relationship underlying any natural process. The notion of causality in physics arises prominently in several circumstances. For relativistic quantum fields, microcausality is the spacelike local commutativity or anticommutativity of fields. A well known example concerning macrocausality is closely related to the Fermi model for propagation of light in quantum electrodynamics~\cite{fermi}. With such a model Fermi proposes the inspection of the radiative processes of two localized quantum systems interacting with a quantum field. These localized systems consist of two two-level atoms separated by some spatial distance $r$. In order to address the issue of causality within this context, Fermi discusses the following gedanken experiment. Consider that, at an arbitrary initial time $\tau_{0}$, atom $1$ is in the excited state, whereas atom $2$ is in the ground state, and both are coupled with a quantum field prepared in the Minkowski vacuum state. Under certain approximations, Fermi argues that the probability for the atom $2$ to be excited (with the field remaining in the vacuum state) remains zero until a time $\tau$ is reached such that $\tau-\tau_{0} \geq r$. We remark that this simple model enables one to investigate causality at the microscopic level.

The Fermi's two-atoms system has been discussed in detail in many works~\cite{heitler,hamilton,fierz,ferretti,milonni1,biswas}. Notwithstanding, the emergence of some questions regarding Fermi's calculation reveals that a more careful consideration of the model could be required. This is important in order to avoid an inaccurate description of the interplay between causal signaling and quantum non-local phenomena. The controversy is justified on the grounds that Fermi's conclusion could be incorrect due to an apparent artificial approximation carried out by him~\cite{Shirokov}, among other issues~\cite{hegerfeldt}. The situation was soon clarified and it was realized that there were no causality problems within the Fermi set up, as long as a proper analysis is undertaken. See for instance Refs.~\cite{buchholz,milonni2,power,kempf}. For more recent discussions and new trends concerning the Fermi problem, we refer the reader the Refs.~\cite{buscemi,zohar,sabin,rey,kempf2,Bouchene,Dick}. 

On the other hand, since the concept of causality is intimately connected with the existence of light cones in the construction of the special relativity, it is not clear how to discuss causality in physical processes when the structure of the light cones is altered. This takes place whenever one allows for the presence of quantum fluctuations of the spacetime. Such quantum effects are expected to arise in any quantum theory of gravity. Hence a comprehensive and thorough account of quantum metric fluctuations would demand a full quantum theory of gravity. Nevertheless, it is still possible to discuss low-energy quantum gravitational effects within an effective-field-theory approach~\cite{donogue,ef1}. On the other hand, one may resort to simplified models which would reproduce the quantum fluctuations of the light cone~\cite{pe9,pe10a,pe10b,pe10c,pe10d,pe10e,pe10f}. 

In this article we propose to investigate how the concept of causality may be disturbed in the presence of light-cone fluctuations. Here we consider that such fluctuations can be treated classically and the effects on the propagation of quantum fields can be described in the context of random differential equations~\cite{pe13,pe15}. We remark that the propagation of acoustic excitations in random media can be envisaged as an analog model for quantum gravity effects~\cite{pe12,prd,ellis}. In turn, within this scenario one may also depart from the inquiry for analog models of quantum gravity effects and conceive the Fermi problem in a more general context of disordered media. To simplify the calculations we assume the units to be such that $\hbar=c=k_{B}=1$.

{\it Photodetection and causality in quantum field theory.}---Let us discuss the Fermi problem in a perturbative framework. Being more specific, we employ a time-dependent perturbation theory (within the interaction picture) in order to evaluate the associated transition probabilities for a finite time interval $[\tau_0,\tau]$. For simplicity and without any loss of generality, we consider two identical static qubits interacting with a massless scalar field. Since the trajectories are stationary, the qubits have stationary states with well defined energy levels. We are working in four dimensional Minkowski spacetime. The total Hamiltonian of the system is given by $H = H_A + H_F + H_I$, where the Hamiltonian of the free atomic system is given by
\begin{equation}
H_{A}=\frac{\omega_0}{2}\,\sum_{j=1}^{2}\sigma_{j}^z,
\label{hd}
\end{equation}
where $\sigma^z_{j} = | e_{j} \rangle\langle e_{j} | - |g_{j} \rangle\langle g_{j} |$, $j = 1, 2$, $|g_1\rangle$ and $|g_2\rangle$ being the ground states of the atoms isolated from each other, with energies $-\omega_0/2$, and $|e_1\rangle$ and $|e_2\rangle$ their respective excited states, with energies $\omega_0/2$. The free Hamiltonian of the massless quantum scalar field is given by
\begin{equation}
H_F = \int \frac{d^3 k}{(2\pi)^3}\, \omega_{{\bf k}}\,a^{\dagger}_{{\bf k}}(t)a_{{\bf k}}(t),
\label{hf}
\end{equation}
where $\omega_{{\bf k}} = |{\bf k}|$. The atom-field interaction Hamiltonian has two local contributions, namely:
\begin{equation}
H_{I} = \lambda\,m_1(\tau)\varphi[x_1(\tau)] + \lambda\,m_2(\tau)\varphi[x_2(\tau)],
\end{equation}
where in the interaction picture one has that $m_{j}(\tau) = e^{i H_{A} \tau} m_{j}(\tau_0) e^{-i H_{A} \tau}$, with $m_{j}(\tau_0) = (i/2)(\sigma_{j}^{-} - \sigma_{j}^{+})$ being the monopole moment operator of the $j$-th atom and $\sigma_{j}^{-} = |g_{j}\rangle \langle e_{j}|$ and $\sigma_{j}^{+} = |e_{j}\rangle \langle g_{j}|$. The time evolution of the system will be described with respect to the proper time $\tau$ of the atoms. We assume that $\lambda \ll 1$. 

In order to access the modification caused by the fluctuating light cone on transition probabilities, we implement a perturbation calculation similar to the one discussed in Refs.~\cite{pe12,prd,ellis}. Let us consider the following random massless scalar Klein-Gordon equation
\begin{equation}
\left[\bigl(1+\,\mu({\bf r})\bigr) \frac{\partial^{2}}{\partial\,t^{2}}
- \Delta\right] \varphi(t,{\bf r})=0,
\label{nami20}
\end{equation}
where $\Delta$ is the three dimensional Laplacian. For $\mu({\bf r}\,)$ we will take a zero-mean Gaussian random function with a correlation function given by
\begin{equation}
\langle \mu({\bf r}\,)\mu({\bf r}\,') \rangle_{\mu} =
\sigma^2\,\delta^{(3)}({\bf r}-{\bf r}\,'),
\label{nami22}
\end{equation}
where $\sigma^2$ quantifies the intensity of random fluctuations. The symbol $\langle\, ...\,\rangle_{\mu}$ denotes an average over all possible realizations of this random variable. Observe that we assume a time-independent random function. Hence, the random Klein-Gordon equation can be solved within a perturbation expansion in the noise function. In addition, note that the Eq.~(\ref{nami20}) is linear in the field variable. This means that the corresponding action of the model is quadratic in the field, resulting in a Gaussian generating functional. Hence all quantum $n$-point Green's functions of the scalar field can be expressed as products of two-point Green's functions. This fact shall be explored in what follows.

As discussed above, the Fermi problem consists in the investigation of causality through a detailed analysis of the energy transfer between a pair of qubits. One assumes that, at the initial time $\tau_{0}$, the system is in the state $|\phi_{i}\rangle = 
|e_1 g_2\rangle \otimes |0_M\rangle$, where $|0_M\rangle$ is the Minkowski vacuum state of the scalar field. Originally, the issue of causality within this setup is examined by setting forth a precise specification of the state of the qubits and also of the quantum scalar field at a later time $\tau$. Notwithstanding, several works have shown the importance of considering more general configurations, see for instance~\cite{power,kempf,Dick}. Hence we will distinguish three possible experimental scenarios in the present survey. The first is similar to the aforementioned circumstance, namely we calculate the probability of finding the system with the atom $2$ in the excited state, the atom $1$ in its ground state {\it and also} the scalar field in the vacuum state $|0_{\mu}\rangle$, at a later time $\tau$. Here $|0_{\mu}\rangle$ is not the usual Minkowski state, but rather a modified vacuum state in the presence of disorder (we also take this to be the initial state of the field). In the second version of the problem, the final state of the field is left unspecified: maintaining the same initial conditions, we evaluate the probability of finding the atom $2$ in the excited state {\it and} the atom $1$ in its ground state. Finally, in the third situation only the state of the atom $2$ is specified at time $\tau$; given the same initial conditions, we discuss the transition probability associated with the transition $g_2 \to e_2$. In this way we intend to provide an answer as complete as possible for the question of how causality (in the precise sense of the Fermi problem) is impacted by (random) light-cone fluctuations in a perturbative framework.

Consider the first situation mentioned above. For a given realization of the disorder, one obtains that the transition probability to the final atom-field state $|\phi_{f}\rangle = |g_1 e_2\rangle\otimes|0_{\mu}\rangle$ is given by (up to fourth order in $\lambda$)
\begin{eqnarray}
{\cal P}_{\phi_{i} \to \phi_{f}}(\mu) &=& \left|\Braket{0_{\mu} e_{2}g_{1}|e_{1}g_{2}0_{\mu}}\right|^{2} 
\nonumber \\ 
&=& \frac{\lambda^{4}}{16}\int_{\tau_{0}}^{\tau}d\tau_{3}\int_{\tau_{0}}^{\tau}d\tau_{4}
\int_{\tau_{0}}^{\tau}d\tau_{1}\int_{\tau_{0}}^{\tau}d\tau_{2} e^{i\omega_{0}(\tau_{3}-\tau_{4})}
\nonumber\\
&\times&\,
e^{-i\omega_{0}(\tau_{1}-\tau_{2})}
G^{*}(\tau_{4},\tau_{3};r) G(\tau_{1},\tau_{2};r),
\label{prob3}
\end{eqnarray}
where $G(x(\tau),x(\tau')) = \langle 0_{\mu} |T\bigl[\varphi(x(\tau))\varphi(x(\tau'))\bigr]| 0_{\mu} \rangle$ is the exact Feynman propagator of the scalar field in the presence of disorder, $r = |{\bf x}_1 - {\bf x}_2|$ being the spatial separation between the qubits and $T$ is the Dyson time-ordering symbol. The Feynman propagator can be written as the following series:
\begin{eqnarray}
\hspace{-0.25cm}
G(x,x') &=& G_{0}(x-x') \nonumber\\
&& +
\sum_{n=1}^{\infty}\int dz_1 \,G_{0}(x-z_1){\cal G}^{(n)}(z_1,x'),
\label{a1}
\end{eqnarray}
where $G_{0}(x-x') = G_{0}(t-t',|{\bf x} - {\bf x}'|)$ is the Feynman propagator for the free field~\cite{birrel}
\begin{eqnarray}
G_{0}(t-t',|{\bf x} - {\bf x}'|) &=& \frac{i}{4\pi^2}\left(\frac{1}{(t-t')^2 - |{\bf x} - {\bf x}'|^2}\right)
\nn\\
&-&\, \frac{1}{8\pi r}\,\frac{(t-t')}{|t-t'|}\Bigl[\delta(|{\bf x} - {\bf x}'| - (t-t')) 
\nn\\
&-&\, \delta(|{\bf x} - {\bf x}'| + (t-t'))\Bigr].
\label{free-feynman}
\end{eqnarray}
The modification in the Feynman propagator due to the presence of disorder is systematized in the following term
\begin{equation}
\hspace{-0.15cm}
{\cal G}^{(n)}(z_1,x') = (-1)^{n}\prod_{j=1}^n L_1(z_j)
\int dz_{j+1} \,G_{0}(z_j,z_{j+1}),
\label{genericterm}
\end{equation}
with $L_1 (x)$ being the random differential operator
\vspace*{-3mm}
\begin{equation}
L_1 (x) = L_1(t,{\bf r\,}) = - \mu({\bf r\,}) \frac{\partial^2}{\partial t^2}.
\label{L1-coord}
\end{equation}
In Eq.~(\ref{genericterm}), it is to be understood that $z_{n+1} = x'$ and
that there is no integration in $z_{n+1}$. Details on the derivations of
the above expressions can be found in Ref.~\cite{prd}.

Now consider the second foregoing case. Up to fourth order in $\lambda$ and for a given realization of the disorder, the transition probability to the final atom-field state $|\psi_{f}\rangle = |g_1 e_2\rangle\otimes|F_{\mu}\rangle$, where 
$|F_{\mu}\rangle$ is a given final state for the scalar field, reads
\begin{eqnarray}
{\cal P}_{\phi_{i} \to \psi_{f}}(\mu) &=& \sum_{F}\left|\Braket{F e_{2}g_{1}|e_{1}g_{2}0_{\mu}}\right|^{2}
\nn\\
&=&\,\frac{\lambda^{4}}{16}\int_{\tau_{0}}^{\tau}d\tau_{3}\int_{\tau_{0}}^{\tau}d\tau_{4}
\int_{\tau_{0}}^{\tau}d\tau_{1}\int_{\tau_{0}}^{\tau}d\tau_{2}\; 
\nn\\
&\times&\,e^{i\omega_{0}(\tau_{3}-\tau_{4})}e^{-i\omega_{0}(\tau_{1}-\tau_{2})}
\nn\\
&\times&\,\Bigl[G^{*}(\tau_{4},\tau_{3};r) G(\tau_{1},\tau_{2};r) 
\nonumber\\
&+&\, G^{+}(\tau_{4},\tau_{1};r) G^{+}(\tau_{3},\tau_{2};r) 
\nn\\
&+&\, G^{+}(\tau_{4},\tau_{2};0) G^{+}(\tau_{3},\tau_{1};0)\Bigr],
\label{prob}
\end{eqnarray}
where completeness relation for the field states was employed. Actually, strictly speaking Eq.~(\ref{prob}) describes the transition probability of the atomic system to the state $|g_1 e_2\rangle$, that is the whole system in any state of the form $|\psi_{f}\rangle$; this is given by a sum over the probabilities for a complete set of field states $|F_{\mu}\rangle$ which is~(\ref{prob}). In addition, notice that the first term inside the brackets produces the same result as in the prior case. Hence when the final state of the quantum field is left unspecified, the transition probability acquires new terms in addition to the precedent contribution.

In the Eq.~(\ref{prob}) $G^{+}(x(\tau),x(\tau'))$ is the exact positive-frequency Wightman function. Since this satisfies the homogeneous wave equation, one could implement a perturbative expansion in the noise field similar to the series expansion derived for the Feynman propagator. One finds
\bea
G^{+}(x,x') &=& G^{+}_{0}(x-x') 
\nn\\
&+&\,\sum_{n=1}^{\infty}\int dz_1 \,G_{\textrm{ret}}(x-z_1){\cal D}^{(n)}(z_1,x'),
\label{a1-wigh}
\eea
where $G^{+}_{0}(x-x') = G^{+}_{0}(t-t',|{\bf x} - {\bf x}'|)$ is the usual positive-frequency Wightman function for the free scalar field:
\begin{equation}
G^{+}_{0}(x-x') = -\frac{1}{4\pi^2}\left[\frac{1}{(t-t'-i\epsilon)^2 - |{\bf x} - {\bf x}'|^2}\right],
\label{free-wigh}
\end{equation}
and $G_{\textrm{ret}}(x-z_1)$ is the retarded Green's function associated with the free Klein-Gordon equation. In addition, the contributions from random fluctuations are encoded in the quantity
\begin{equation}
\hspace{-0.15cm}
{\cal D}^{(n)}(z_1,x') = (-1)^{n}\prod_{j=1}^n L_1(z_j)
\int dz_{j+1} \,G^{+}_{0}(z_j,z_{j+1}),
\label{genericterm-wigh}
\vspace{1mm}
\end{equation}
with $L_1 (x)$ being the random differential operator defined in Eq.~(\ref{L1-coord}).

Finally, in the third and last situation, we wish to calculate the transition probability to the final atom-field state 
$|\Phi_{f}\rangle = |S e_2\rangle\otimes|F_{\mu}\rangle$, where $|F_{\mu}\rangle$ is as above and $|S\rangle$ is any given state of the atom $1$. That is, the probability to find the atom $2$ in the state $|e_2\rangle$, in other words the whole system in any state of the form $|\Phi_{f}\rangle$. This is obtained by a double sum over the probabilities for a complete set of field states 
$|F_{\mu}\rangle$ and for a complete set of atomic states $|S\rangle$ of atom $1$. Up to fourth order in $\lambda$ and for a given realization of the disorder, one has that
\bea
{\cal P}_{\phi_{i} \to \Phi_{f}}(\mu)&=&\sum_{F,S}|\Braket{Fe_{2}S|e_{1}g_{2}0_{\mu}}|^{2} 
= {\cal P}(\mu) 
\nn\\
&+&\, {\cal P}_{\phi_{i} \to \psi_{f}}(\mu) + \Delta{\cal P}(\mu),
\label{prob2}
\eea
where
\begin{widetext}
\begin{eqnarray}
\Delta{\cal P}(\mu) &=& 
-\frac{\lambda^4}{16} \int_{\tau_{0}}^{\tau}d\tau_1 \int_{\tau_{0}}^{\tau}d\tau_2 \int_{\tau_{0}}^{\tau}d\tau_{3}\int_{\tau_{0}}^{\tau}d\tau_{4}\,
e^{i\omega_{0}(\tau_{3}-\tau_{4})} e^{-i\omega_{0}(\tau_{1}-\tau_{2})}
\nonumber\\
&\times&\,\biggl\{G^{+}(\tau_1,\tau_2;r) G^{+}(\tau_3,\tau_4;r)
[\theta(\tau_2-\tau_4)\theta(\tau_3-\tau_4) + \theta(\tau_1-\tau_3)\theta(\tau_4-\tau_3)]
\nonumber\\
&+&\,G^{+}(\tau_1,\tau_2;r) G^{+}(\tau_4,\tau_3;r)
[\theta(\tau_2-\tau_4)\theta(\tau_4-\tau_3) + \theta(\tau_1-\tau_3)\theta(\tau_3-\tau_4)]
\nonumber\\
&+&\,G^{+}(\tau_1,\tau_4;r) G^{+}(\tau_2,\tau_3;r)
\Bigl[\bigl(\theta(\tau_2-\tau_4)+\theta(\tau_3-\tau_4)\bigr)\theta(\tau_2-\tau_3)
+ \bigl(\theta(\tau_4-\tau_2)+\theta(\tau_1-\tau_2)\bigr)\theta(\tau_4-\tau_1)\Bigr]\biggr\}
\nonumber\\
&-&\,\frac{\lambda^4}{16} \int_{\tau_{0}}^{\tau}d\tau_1 \int_{\tau_{0}}^{\tau}d\tau_2 \int_{\tau_{0}}^{\tau}d\tau_{3}\int_{\tau_{0}}^{\tau}d\tau_{4}\,
\Bigl\{e^{i\omega_{0}(\tau_{3}-\tau_{4})} e^{-i\omega_{0}(\tau_{1}-\tau_{2})}
\Bigl[G^{+}(\tau_1,\tau_3;0) G^{+}(\tau_2,\tau_4;0)\bigl(1+\theta(\tau_4-\tau_3)\bigr)
\nonumber\\
&+&\, G^{+}(\tau_1,\tau_2;0) G(\tau_3,\tau_4;0)
+ G^{+}(\tau_1,\tau_4;0) G^{+}(\tau_2,\tau_3;0)\theta(\tau_4-\tau_3)\Bigr]\theta(\tau_2-\tau_4)\theta(\tau_2-\tau_3)
+ \textrm{c.c.}\Bigr\}. 
\label{delta-prob}
\end{eqnarray}
\end{widetext}
In order to obtain such results, we used the completeness relation for the atomic and the field states. In Eq.~(\ref{prob2}), 
${\cal P}(\mu)$ is given by 
\beq
{\cal P}(\mu) = \frac{\lambda^{2}}{4} \int_{\tau_{0}}^{\tau}d\tau'\int_{\tau_{0}}^{\tau}d\tau''
e^{i\omega_{0}(\tau''-\tau')}
G^{+}(\tau',\tau'';0).
\label{prob-sup}
\eeq
whereas ${\cal P}_{\phi_{i} \to \psi_{f}}(\mu)$ is given by Eq.~(\ref{prob}).~The $r$-independent terms in Eq.~(\ref{prob2}) describe the contributions to the transition probability that arise from the interaction of each atom with the field vacuum alone. Concerning our current analysis of causality effects, the relevant terms that actually comprise the description of the impact of atom $1$ on atom $2$ are given by all $r$-dependent terms in the Eqs.~(\ref{prob}) and~(\ref{delta-prob}). These are the terms that we will duly consider in the present investigations. A discussion on which terms are relevant in order to assess the impact of atom $1$ on atom $2$ can also be found in Refs.~\cite{kempf,kempf2,jon}.

These last results lead to the following conclusion. Suppose the existence of terms in the transition probability with $\Delta\tau = \tau - \tau_0 < r$, where $\Delta\tau$ is the observational time interval. This signalises an explicit violation of causality. Yet, whenever the Wightman function is of the form
\beq
G^{+}(t-t';r) = -\frac{1}{4\pi^2}\frac{f(t-t',r)}{[(t-t'-i\epsilon)^2 - r^2]^{n}},
\label{cond1}
\eeq
and the portion of the Feynman propagator that generates acausal effects (as discussed in the next section) is also of a similar form, namely
\beq
G_{\textrm{NC}}(t-t';r) = \frac{i}{4\pi^2}\frac{f(t-t',r)}{[(t-t')^2 - r^2]^{n}},
\label{cond2}
\eeq
the noncausal terms cancel each other in Eq.~(\ref{prob2}) and the result is strictly causal (for $\Delta\tau < r$ the limit $\epsilon \to 0$ can be taken directly in the Wightman function). In Eqs.~(\ref{cond1}) and~(\ref{cond2}) $n$ is a non-negative integer and $f(t-t',r)$ is an even function of $t-t'$.

{\it Unperturbed contributions.}---Let us first discuss the terms that would arise even in the absence of random fluctuations of the light cone. Consider the situation in which the final state of the system is given by $|\phi_{f}\rangle$. Introducing the variables $\xi = \tau' - \tau''$ and $\eta = \tau' + \tau''$, expression~(\ref{prob3}) becomes
\begin{equation}
{\cal P}^{(0)}_{\phi_{i} \to \phi_{f}} = \frac{\lambda^{4}}{16}
\Bigg|\int_{-\Delta\tau}^{\Delta\tau}d\xi \,(\Delta\tau - |\xi|)\, e^{-i\omega_0 \xi}\,G_{0}(\xi, r)\Bigg|^2.
\label{amplitude2}
\end{equation}
where $G_{0}(t-t';r)$ is given by Eq.~(\ref{free-feynman}). When one inserts the expression for the free Feynman propagator in Eq.~(\ref{amplitude2}), one gets two contributions for each of the double time integrals. The contribution related with the Dirac delta function will produce a Heaviside theta function $\theta(-r + \Delta\tau)$. This term is in accordance with the definition of causality given by Ref.~\cite{stu}. On the other hand, the other contribution yields a finite term for $\Delta\tau < r$. The latter is a function proportional to $1/(\omega_0^2 r^2)$ [in fact, one can show that it is a combination of the trigonometric integrals $\Ci(\omega_0 r), \Si(\omega_0 r)$ and trigonometric functions $\cos(\omega_0 r), \sin(\omega_0 r)$ times $1/(\omega_0^2 r^2)$]. As previously outlined, this outcome has been discussed in the literature in many papers: The dispute is on the causal behavior of the transition probability. However, as argued by Pauli, Stueckelberg and others~\cite{stu,pauli}, the concept of causality in relativistic quantum field theory has meaning only in the wave zone $\omega_0 r \gg 1$. Hence since the anomalous term mentioned above vanishes in the wave zone, causality as defined above is preserved.

In the second case outlined above, the final state of the system is given by $|\psi_{f}\rangle$, i.e., only the final states of the qubits are specified. Employing the same change of variables as above and after using the usual expression for the free postive-frequency Wightman function $G^{+}_{0}(t-t';r)$, we obtain three kinds of terms. One is identical to the previous result which was discussed above, given by Eq.~(\ref{amplitude2}). The second term in the brackets of Eq.~(\ref{prob}) yields a finite contribution for $\Delta\tau < r$ which is proportional to $g(\omega_0,r,\Delta t)/(\omega_0^2 r^2)$, where $g(\omega_0,r,\Delta t)$ can be expressed in terms of trigonometric integrals and trigonometric functions. The last term inside the brackets in Eq.~(\ref{prob}) produces a term that reproduces identical outcomes discussed in Ref.~\cite{BNSvaiter}. In particular, such a contribution does not depend on $r$ and hence it does not play a part in the analysis of the influence of atom $1$ upon the radiative processes of atom $2$. In conclusion, one says that within this setting causality is preserved in the process of energy transfer between the atoms, as prescribed above (i.e., only in the wave zone). 

Finally, let us discuss the situation in which only the final state of the atom $2$ is specified. The insertion of the free Green's functions of our problem in Eq.~(\ref{prob2}) yields three terms, two of which have the same form and behavior as the previous case. The last term is related with the contribution $\Delta{\cal P}(\mu)$. Since conditions~(\ref{cond1}) and~(\ref{cond2}) are met by the Wightman function and the Feynman propagator of the free scalar field, one can easily see that noncausal terms are found to cancel each other and as such the transition probability is strictly causal when the measurement is inclusive, recovering the results of Ref.~\cite{power}.

{\it Light-cone fluctuations and causality.}---Following the aforementioned discussion, the purpose of this work is to discuss the notion of causality in disordered media in the framework of the well known Fermi's problem of two-qubits system. Here we consider all terms up to second order in $\mu$ in the perturbative series for the Feynman propagator and the Wightman function. In addition, we keep terms up to order $\sigma^2$ in all of our calculations after considering the random averages.

Consider the transition to the final state $|\phi_{f}\rangle$. After performing the averages over the noise function, one obtains the following probability, 
\begin{eqnarray}
\langle {\cal P}_{\phi_{i} \to \phi_{f}}(\mu) \rangle_{\mu}  &=&  \frac{\lambda^{4}}{16}
\int_{\tau_{0}}^{\tau}d\tau_{3}\int_{\tau_{0}}^{\tau}d\tau_{4}
\int_{\tau_{0}}^{\tau}d\tau_{1}\int_{\tau_{0}}^{\tau}d\tau_{2} 
\nn\\
&\times&\,e^{i\omega_{0}(\tau_{3}-\tau_{4})} e^{-i\omega_{0}(\tau_{1}-\tau_{2})}
\nn\\
&\times&\,\Bigl[G^{*}_{0}(\tau_4-\tau_3;r)G_{0}(\tau_1-\tau_2;r)
\nn\\
&+&\, G^{*}_{0}(\tau_4-\tau_3;r)I(\tau_1-\tau_2;r) 
\nn\\
&+&\, G_{0}(\tau_1-\tau_2;r) I^{*}(\tau_4-\tau_3;r)\Bigr].
\label{prob-media1}
\end{eqnarray}
The function $I(\Delta t,|\Delta{\bf x}|)$ (or rather its ``finite-time Fourier transform") is the basic quantity which ultimately accounts for possible causality-violation effects. It is given by
\bea
I(\Delta t,|\Delta{\bf x}|) &=& \frac{6 i\sigma^2}{(2\pi)^3} \,\left[ \frac{{\cal F}(\Delta t, |\Delta {\bf x}|)\, \theta(\Delta t)}
{\left[(\Delta t-i\epsilon)^2 - |\Delta{\bf x}|^2\right]^5} \right.
\nn\\
&+&\, \left. \frac{{\cal F}(-\Delta t, |\Delta {\bf x}|)\, \theta(-\Delta t)}{\left[(\Delta t+i\epsilon)^2 
- (\Delta{\bf x})^2\right]^5}\right],
\eea
with $\Delta t=t-t'$, $\Delta{\bf x}={\bf x}-{\bf x}'$ and
\bea
\hspace{-1mm}
{\cal F}(\Delta t, |\Delta {\bf x}|) &=&  \Delta t\Bigl[5(\Delta t - i\epsilon)^4  +  10 (\Delta t - i\epsilon)^2 |\Delta {\bf x}|^2
\nn\\
&+&\, |\Delta {\bf x}|^4\Bigr] - 4(\Delta t - i\epsilon)\,\Bigl[(\Delta t - i\epsilon)^4 - |\Delta {\bf x}|^4 \Bigr].
\nn\\
\eea
The first term in the brackets is simply the contribution discussed above. The double time integrals of $I$ (which are just the leading-order disorder correction to the transition amplitudes) produce a finite term for $\Delta\tau < r$ which does not vanish in the wave zone $\omega_0 r \gg 1$:
\bea
{\cal I}_{\phi_{i} \to \phi_{f}}
&=&\,\int_{\tau_{0}}^{\tau}d\tau_{1}\int_{\tau_{0}}^{\tau}d\tau_{2} 
e^{-i\omega_{0}(\tau_{1}-\tau_{2})}\,I(\tau_1-\tau_2;r)
\nn\\
&\xrightarrow{\omega_0 r \gg 1}& -\frac{i\pi\sigma^2}{(2\pi)^4}\omega_0^3\,\sin(\omega_0\Delta\tau),\,\,\,\Delta\tau < r.
\label{viol}
\eea
This result clearly shows a violation of causality even within the wavezone. One possible interpretation is that the arguments previously invoked on the definition of causality simply do not apply in the presence of light-cone fluctuations. However by considering transition probabilities instead of transition amplitudes, one notices that ${\cal I}_{\phi_{i} \to \phi_{f}}$ gets multiplied by a term coming from $G_{0}$ that vanishes in the wave zone. Hence causality as perceived by Pauli and Stueckelberg is preserved as long as one only considers the transition probabilities. 

Concerning the transition to the final state $|\psi_{f}\rangle$, one finds, after performing the random averages
\vspace{-1mm}
\begin{widetext}
\begin{eqnarray}
\langle {\cal P}_{\phi_{i} \to \psi_{f}}(\mu) \rangle_{\mu} &=& \frac{\lambda^{4}}{16}\int_{\tau_{0}}^{\tau}d\tau_{3}\int_{\tau_{0}}^{\tau}d\tau_{4}
\int_{\tau_{0}}^{\tau}d\tau_{1}\int_{\tau_{0}}^{\tau}d\tau_{2} e^{i\omega_{0}(\tau_{3}-\tau_{4})}
e^{-i\omega_{0}(\tau_{1}-\tau_{2})}
\Bigl[G^{*}_{0}(\tau_4-\tau_3;r)G_{0}(\tau_1-\tau_2;r)
\nonumber\\
&+&\, G^{+}_{0}(\tau_4-\tau_1;r)G^{+}_{0}(\tau_3-\tau_2;r)
+ G^{+}_{0}(\tau_4-\tau_2;0)G^{+}_{0}(\tau_3-\tau_1;0)
\nn\\
&+&\,G^{*}_{0}(\tau_4-\tau_3;r)I(\tau_1-\tau_2;r) + G_{0}(\tau_1-\tau_2;r) I^{*}(\tau_4-\tau_3;r)
\nn\\
&+&\,G^{+}_{0}(\tau_4-\tau_1;r)I^{+}(\tau_3-\tau_2;r) + G^{+}_{0}(\tau_3-\tau_2;r)I^{+}(\tau_4-\tau_1;r)
\nn\\
&+&\,G^{+}_{0}(\tau_4-\tau_2;0)I^{+}(\tau_3-\tau_1;0) + G^{+}_{0}(\tau_3-\tau_1;r)I^{+}(\tau_4-\tau_2;0)\Bigr]
\label{prob-media2}
\end{eqnarray}
\end{widetext}
where $I^{+}(t-t',r) = I(t-t',r)|_{t-t' > 0}$. The zeroth-order contributions in the noise field and the terms related with the function $I(t-t';r)$ have been examined above. On the other hand, a straightforward analysis unveils that the double time integrals over the function $I^{+}(t-t';r)$ vanish for $\Delta\tau < r$. In turn, the last terms inside the brackets of  Eq.~(\ref{prob-media2}) do not depend on the distance between the qubits. Hence causality is preserved for this transition in the presence of weak disorder, but only in the wave zone. 

The last situation that we study in this work considers the transition to the final state $|\Phi_{f}\rangle$. After performing the random averages, one obtains
\bea
\langle {\cal P}_{\phi_{i} \to \Phi_{f}}(\mu)\rangle_{\mu} &=& \langle{\cal P}(\mu)\rangle_{\mu} 
\nn\\
&+&\, \langle {\cal P}_{\phi_{i} \to \psi_{f}}(\mu)\rangle_{\mu} + \langle\Delta{\cal P}(\mu)\rangle_{\mu},
\label{prob-media3}
\eea
As previously discussed, we are considering only the $r$-dependent terms from the above equation, since these are precisely the contributions that represent the impact of atom $1$ on atom $2$ with respect to the causality effects addressed in this work.

The first term of Eq.~(\ref{prob-media3}) is $r$-independent whereas the second term arises in the transition probability to the final state $|\psi_{f}\rangle$. The contribution coming from $\langle\Delta{\cal P}(\mu)\rangle_{\mu}$ is more involved. It is given by quadruple time integrals of terms such as
$
G^{+}_{0}(\tau_1-\tau_2;r)I^{+}(\tau_3-\tau_4;r) + G^{+}_{0}(\tau_3-\tau_4;r)I^{+}(\tau_1-\tau_2;r).
$
Such integrals engender cumbersome expressions which are hard to grasp in a simple way. However, the analysis of causality follows almost straightforward from the arguments presented above. It is easy to see that the function $I(t-t';r)$ does not satisfy the conditions~(\ref{cond1}) and~(\ref{cond2}). Hence in the presence of disorder the probability is not strictly causal when the measurement is inclusive. Notwithstanding, such integrals will produce contributions that are proportional to 
$h(\omega_0,r,\Delta t)/(\omega_0^m r^m)$, where $m$ is a positive integer and, similar to the previous cases, $h(\omega_0,r,\Delta t)$ can be written in terms of trigonometric integrals and trigonometric functions. Such contributions are suppressed by powers of $\omega_0 r$. Hence, if one insists in maintaining causality, one must make use of the arguments of Pauli and Stueckelberg and consider the limit $\omega_0 r \gg 1$. In this way, the terms for $\Delta\tau < r$ vanish and causality is restored.

{\it Conclusions.}---There is a contentious discussion on the causal behavior regarding the radiative processes involving two identical qubits coupled with a quantum field. In this article we proposed to study the issue of the macrocausality for the case in which we allow for random fluctuations of the light cones. We have considered three distinct configurations, each of which portraying a different specification of the system at the time of measurement. In the case of a Minkowski spacetime with fixed light cones, we recovered the usual result that states that the macrocausality is maintained, at least up to fourth-order in time-dependent perturbation theory. In particular, when the final measurement is exclusively made on the atom $2$, we found that the energy transfer between the atoms is exactly causal, that is, the transition probability does not comprise terms with $\Delta \tau < r$. In fact Ref.~\cite{kempf} demonstrates that this result holds to all orders in perturbation theory. On the other hand, when one considers the system in a disordered environment, one finds that causality has a well-defined meaning solely in the wave zone, in all the situations studied above. In other words, by using the reasoning developed by Pauli the usual concept of causality in the presence of random fluctuations of the light cone makes sense only when $\omega_0 r \gg 1$, in the limit of weak disorder. We conjecture that this outcome can be extended to all orders in perturbation theory.

The typical distance $r_{0}$ outside of the wave zone within which the violation of causality can be strongly upheld may be estimated with the following argument. For situations such that $\sigma^{2}\omega^{3}_{0} \approx \omega_{0} r_{0}$, one finds that $r_{0} \approx \sigma^{2}\omega^{2}_{0}$. This gives a relation between $r_{0}$ and the intensity of random fluctuations for a given atomic energy gap. For distances such that $\omega_{0} r \gg \omega_{0} r_{0}$, the disorder effects are weak and a perturbative series expansion in $\mu$ is thus warrant; nonetheless violation of causality is at any rate verified. In the wave zone $\omega_{0} r \gg 1$, random fluctuations become negligible and hence causality is retrieved.

In the context of the present investigations, one could argue that the violation of strict causality is connected with the unbounded nature of the Gaussian noise employed here. Indeed, it has been shown that domains in which $1+\,\mu({\bf r}) < 0$ lead to unstable wave modes and consenquently the emergence of sharply growing contributions to the density of states at high frequencies~\cite{altland}. This in turn could be interpreted as an indication that Gaussian noise (or more generally unbounded distributions) could blur the usual notion of causality. However, the conservation of causality in the wave zone could be an indication that the Pauli-Jordan function (which is the commutator of two scalar field operators) may still have a causal support, at least in a broader sense. On the other hand, as discussed in the text the violation of strict causality is connected with the parity behavior of the Wightman function and the Feynman propagator as functions of $\Delta\tau$ for $\Delta\tau < r$. Inspection of the perturbative expansion for both of these Green's functions and simple use of Fourier transforms show that there is a strong evidence that a delta-correlated disorder in both space and time could drastically alter the conclusions of this work. Namely, in this case we expect such Green's functions with uncorrelated random fluctuations in space-time to comply with the conditions~(\ref{cond1}) and~(\ref{cond2}) and in this way strict causality could be warranted when the measurement is inclusive. 

The discussion of the latter paragraph implies that violation of strict causality in our study might be restricted to the class of uncorrelated noises in space. More general situations, still within unbounded distributions, produce distinct upshots. In turn, it is not clear {\it a priori} how bounded fluctutations of the light cone could lead to similar violations of causality. It is known that the smearing of the light cone modifies virtual processes; since wave-zone effects are manifestations of real photons whereas near-field processes are a result of a collection of real and virtual photons effects, one could expect substantial changes in the orthodox picture. This is an important question but lies outside of the scope of the present work. This issue will be reserved for future publications.

An invaluable extension of our results would be to ascertain the above conclusions for the case of quantum fluctuations of the light cone. It would be of particular interest to confront the conclusions and interpretations in both situations. An akin situation would be to consider the role of a spacetime foam in the radiative processes of the Fermi two-atoms problem~\cite{garay}. It is known that restrictions on the measurability of spacetime distances seem to emerge quite naturally in a framework which properly takes into account both quantum mechanical and general-relativistic effects~\cite{aa1}. This is expected to modify our current understanding on the concept of causality within Fermi's set up. Such subjects are under investigation by the authors.
 
This paper was partially supported by Conselho Nacional de Desenvolvimento Cient\'ifico e Tecnol{\'o}gico (CNPq, Brazil).


\vspace{-5mm}
\begin{thebibliography}{99}
%
\bibitem{fermi} E. Fermi, Rev. Mod. Phys. {\bf 4}, 87 (1932).
%
\bibitem{heitler} W. Heitler and S. T. Ma, Proc. R. Ir. Acad. {\bf 52}, 123 (1949).
%
\bibitem{hamilton} J. Hamilton, Proc. Phys. Soc. A {\bf 62}, 12 (1949).
%
\bibitem{fierz} M. Fierz, Helv. Phys. Acta {\bf 23}, 731 (1950).
%
\bibitem{ferretti} B. Ferretti, in {\it Old and New Problems in Elementary Particles},
edited by G. Puppi (Academic Press, New York,1968), p. 108.
%
\bibitem{milonni1} P. W. Milonni and P. L. Knight, Phys. Rev. A {\bf 10}, 1096 (1974).
%
\bibitem{biswas} A. K. Biswas, G. Compagno, G. M. Palma, R. Passante and F. Persico, Phys. Rev. A {\bf 42}, 4291 (1990);
Phys. Rev. A {\bf 44}, 798 (1991).
%
\bibitem{Shirokov} M.I. Shirokov, Yad. Fiz. {\bf 4}, 1077 (1966).
%
\bibitem{hegerfeldt} G. C. Hegerfeldt, Phys. Rev. Lett. {\bf 72}, 596 (1994).
%
\bibitem{buchholz} D. Buchholz and J. Yngvason, Phys. Rev. Lett. {\bf 73}, 613 (1994).
%
\bibitem{milonni2} P. W. Milonni, D. F. V. James and H. Fearn, Phys. Rev. A {\bf 52}, 1525 (1995).
%
\bibitem{power}  E. A. Power and T. Thirunamachandran, Phys. Rev. A {\bf 56}, 3395 (1997).
%
\bibitem{kempf} M. Cliche and A. Kempf, Phys. Rev. A {\bf 81}, 012330 (2010).
%
\bibitem{buscemi} F. Buscemi and G. Compagno, Phys. Rev. A {\bf 80}, 022117 (2009)
%
\bibitem{zohar} E. Zohar and B. Reznik, N. Journ. Phys. {\bf 13}, 075016 (2011).
%
\bibitem{sabin} C. Sabin, M. delRey, J. J. Garcia-Ripoll and J. Leon, Phys. Rev. Lett. {\bf 107}, 150402 (2011).
%
\bibitem{rey} M. del Rey, C. Sab\'in and J. L\'eon, Phys. Rev. A {\bf 85}, 045802 (2012).
%
\bibitem{kempf2} R. H. Jonsson, E. Mart\'in-Mart\'inez and A. Kempf, Phys. Rev. A {\bf 89}, 022330 (2014).
%
\bibitem{Bouchene} S. Derouault and M. A. Bouchene, Phys. Rev. A {\bf 90}, 023828 (2014).
%
\bibitem{Dick} R. Dickinson, J. Forshaw and P. Millington, Phys. Rev. D {\bf 93}, 065054 (2016).
%
\bibitem{donogue} John F. Donoghue,  Phys. Rev. Lett. {\bf 72}, 2996 (1994), ibid, Phys. Rev. D {\bf 50}, 3874 (1994).
%
\bibitem{ef1} W. Dittrich and M. Reuter, {\em{Effective Lagrangians in Quantum Electrodynamics}}, Springer-Verlag, Berlin (1986).
%
\bibitem{pe9} L. H. Ford, Phys. Rev. D {\bf 51}, 1692 (1995).
%
\bibitem{pe10a} L. H. Ford and N. F. Svaiter, Phys. Rev. D {\bf 56}, 2226 (1997).
%
\bibitem{pe10b} L. H. Ford and N. F. Svaiter, Phys. Rev. D {\bf 54}, 2640 (1996).
%
\bibitem{pe10c} H. Yu and L. H. Ford,  Phys. Rev. D {\bf 60}, 084023 (1999).
%
\bibitem{pe10d} R. T. Thompson and L. H. Ford,  Phys. Rev. D {\bf 78}, 024014 (2008).
%
\bibitem{pe10e} R. T. Thompson and L. H. Ford, Class. Quant. Grav. {\bf 25}, 154006 (2008).
%
\bibitem{pe10f} H. Yu, N. F. Svaiter and L. H. Ford, Phys. Rev. D {\bf 80}, 124019 (2009).
%
\bibitem{pe13} A. Ishimaru, {\it Wave Propagation and Scattering in Random Media} (Academic, New York, 1978).
%
\bibitem{pe15} J. Pierre Fouque, J. Garnier, G. Papanicolaou and K. Solna, {\it Wave Propagation and Time Reversal in Randomly
Layered Media} (SpingerScience+Bussiness Media, LLC, 2007).
%
\bibitem{pe12} G. Krein, G. Menezes and N. F. Svaiter, Phys. Rev. Lett. {\bf 105}, 131301 (2010).
%
\bibitem{prd} E. Arias, E. Goulart, G. Krein, G. Menezes and N. F. Svaiter, Phys. Rev. D {\bf 83}, 125022 (2011).
%
\bibitem{ellis} G. Krein, G. Menezes, E. Arias and N. F. Svaiter, Int. Jour. Mod. Phys. A {\bf 27}, 1250129 (2012).
%
\bibitem{birrel} N. D. Birrell and P. C. W. Davis, {\it Quantum Fields in Curved Space} (Cambridge University Press, New York, 1982).
%
\bibitem{jon} R. H. Jonsson, J. Phys. A: Math. Theor. {\bf 50}, 355401 (2017).
%
\bibitem{stu} E. C. G. Sueckelberg and D. Rivier, Helv. Phys. Acta {\bf 23}, 215 (1950).
%
\bibitem{pauli} W. Pauli, {\it Pauli lectures on physics: Selected Topics in Field Quantization} (Dover Publications, New York, 1973).
%
\bibitem{BNSvaiter} B. F. Svaiter e N. F. Svaiter, Phys. Rev. D {\bf 46}, 5267 (1992).
%
\bibitem{altland} V. Gurarie and A. Altland, Phys. Rev. Lett. {\bf 94}, 245502 (2005).
%
\bibitem{garay} L. J. Garay, Phys. Rev. Lett. {\bf 80}, 2508, (1998); ibid. Phys. Rev. D {\bf 58}, 124015 (1998).
%
\bibitem{aa1} G. Amelino-Camelia, Nature, {\bf 398}, 216 (1999); ibid. Mod.Phys. Lett. A {\bf 9}, 3415 (1994).
%

\end{thebibliography}
\end{document}